\documentclass[%
 aip,
 amsmath,amssymb,
 reprint,%
]{revtex4-1}

\usepackage{graphicx}
\usepackage{dcolumn}
\usepackage{bm}

\usepackage[utf8]{inputenc}
\usepackage[T1]{fontenc}
\usepackage{mathptmx}
\usepackage{etoolbox}

\makeatletter
\def\@email#1#2{%
 \endgroup
 \patchcmd{\titleblock@produce}
  {\frontmatter@RRAPformat}
  {\frontmatter@RRAPformat{\produce@RRAP{*#1\href{mailto:#2}{#2}}}\frontmatter@RRAPformat}
  {}{}
}%
\makeatother
\begin{document}

\preprint{AIP/123-QED}

\title[Zero-field spin wave turns]{Zero-field spin wave turns}
\author{Jan Kl\'{i}ma}
\thanks{Authors to whom correspondence should be addressed: [Jan Kl\'{i}ma, jan.klima4@vutbr.cz; Michal Urb\'{a}nek, michal.urbanek@ceitec.vutbr.cz]}
\affiliation{Faculty of Mechanical Engineering, Institute of Physical Engineering, Brno University of Technology, Technick\'{a} 2, Brno, 616 69, Czech Republic}
\author{Ond\v{r}ej Wojewoda}
\affiliation{%
 CEITEC BUT, Brno University of Technology, Purky\v{n}ova 123, Brno, 612 00, Czech Republic
}%
\author{V\'{a}clav Rou\v{c}ka}
\affiliation{Faculty of Mechanical Engineering, Institute of Physical Engineering, Brno University of Technology, Technick\'{a} 2, Brno, 616 69, Czech Republic}
\author{Tom\'{a}\v{s} Moln\'{a}r}
\affiliation{%
 CEITEC BUT, Brno University of Technology, Purky\v{n}ova 123, Brno, 612 00, Czech Republic
}
\author{Jakub Holobr\'{a}dek}
\affiliation{%
 CEITEC BUT, Brno University of Technology, Purky\v{n}ova 123, Brno, 612 00, Czech Republic
}%
\author{Michal Urb\'{a}nek}
\thanks{Authors to whom correspondence should be addressed: [Jan Kl\'{i}ma, jan.klima4@vutbr.cz; Michal Urb\'{a}nek, michal.urbanek@ceitec.vutbr.cz]}
\affiliation{Faculty of Mechanical Engineering, Institute of Physical Engineering, Brno University of Technology, Technick\'{a} 2, Brno, 616 69, Czech Republic}
\affiliation{%
 CEITEC BUT, Brno University of Technology, Purky\v{n}ova 123, Brno, 612 00, Czech Republic
}%
\date{\today}

\begin{abstract}
Spin-wave computing, a potential successor to CMOS-based technologies, relies on the efficient manipulation of spin waves for information processing. While basic logic devices like magnon transistors, gates, and adders have been experimentally demonstrated, the challenge for complex magnonic circuits lies in steering spin waves through sharp turns. In this study we demonstrate with micromagnetic simulations and Brillouin light scattering microscopy experiments, that dipolar spin waves can propagate through 90-degree turns without distortion. The key lies in carefully designed in-plane magnetization landscapes, addressing challenges posed by anisotropic dispersion. The experimental realization of the required magnetization landscape is enabled by spatial manipulation of the uniaxial anisotropy using corrugated magnonic waveguides. The findings presented in this work should be considered in any magnonic circuit design dealing with anisotropic dispersion and spin wave turns.
\end{abstract}

\maketitle

Over the last decade, spin-wave computing has been intensely investigated as a promising candidate to complement and surpass CMOS-based technologies\cite{chumak22}. Many of the basic logic devices, such as a magnon transistor\cite{chumak14} or various types of gates\cite{schneider08} and adders\cite{wang20} have already been demonstrated experimentally. They make use of spin waves, or their quasiparticles – magnons, as energy-efficient information carriers, which can perform both Boolean and wave-based operations\cite{kiechle23}. However, to be able to combine these devices into a complex magnonic circuit, the steering of spin waves needs to be solved. To build complex magnonic networks, waveguides operating at zero external magnetic field\cite{nikolaev23, haldar16, haldar17, flajsman20, turcan21} and with sharp bends (ideally 90° or more) must be realized. So far, the experimentally realized waveguide bends\cite{vogt14, schneider08, wang20, albisetti18} were typically very gradual and they rarely turned more than 45°. The challenge, when passing spin waves through a turn in a magnonic waveguide, lies in maintaining the wave mode and in preventing unwanted reflections from the waveguide edges\cite{xing13}.

These problems are caused by anisotropic dispersion relation of dipolar spin waves in in-plane magnetized structures, where the change in the spin wave direction leads to a dispersion mismatch with detrimental result on the spin wave propagation. The situation should be better for spin waves with isotropic dispersion, e.g., for forward-volume spin waves in out-of-plane magnetized waveguides and for short-wavelength, exchange-dominated spin waves. However, these spin waves have either short decay lengths (forward volume dipolar spin waves) or are experimentally very difficult to excite and detect (exchange-dominated spin waves). This means that turns of these types of spin waves could be studied so far only by micromagnetic simulations\cite{mieszczak20, xing13}.

Here we show that in appropriately designed in-plane magnetization landscape, dipolar spin waves can be smoothly steered through a 90° turn without any distortion. First, we demonstrate the concept of smooth steering of dipolar spin waves with micromagnetic simulations, and later we show with phase-resolved Brillouin light scattering (BLS) microscopy experiments that also the experimental realization is feasible.

\begin{figure}
\includegraphics[width=85mm]{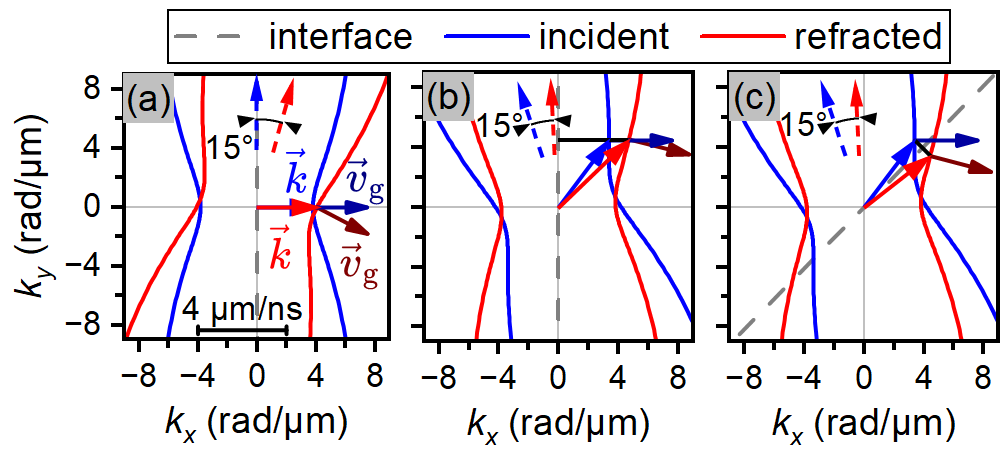}
\caption{\label{fig:dispersions} \textit{k}-space representation of spin wave refraction on an interface between two regions with magnetization rotated by 15° (indicated by dashed arrows). Isofrequency curves ($f = 6\,$GHz) of the spin wave dispersions before (after) refraction are shown as red (blue) curves. The interface between the regions is shown as gray dashed line. (a)~The magnetization in the first region is parallel to the interface and the \textit{k}-vector is perpendicular to the interface. (b)~The magnetization in the first region is rotated by 108°, the \textit{k}-vector by 53°, and the interface by 90° from the $x$-axis. (c)~The magnetization in the first region is rotated by 108°, the \textit{k}-vector by 53°, and the interface by 45.5° from the $x$-axis. The values used for the calculation were\cite{turcan21}: film thickness $t = 10\,$nm, saturation magnetization $M_{\mathrm{s}} = 830\,$kA/m, uniaxial magnetic anisotropy $K_{\mathrm{u}} = 8.3 \,\mathrm{kJ/m^3}$, gyromagnetic ratio $\gamma/2\pi = 29.3\,$GHz/T, and exchange constant $A_{\mathrm{ex}} = 16\,$pJ/m.}.
\end{figure}

When a spin wave refracts at an interface with totally unpinned spins, the component of \textit{k}-vector parallel to the interface is preserved\cite{nikitov01}. In the perpendicular direction the translational symmetry is broken, and the corresponding \textit{k}-vector component can change\cite{stigloher16, stigloher18}. In Fig.~\ref{fig:dispersions}(a), we can see the situation where the spin wave propagates from the left region towards the interface in $\vec{k}\perp \vec{M}$ geometry [Damon-Eschbach (DE)]. As the wave arrives at the interface and refracts into the region with rotated magnetization, it accommodates its \textit{k}-vector to the new dispersion relation (red isofrequency curve). To obey the new dispersion and to conserve the component parallel to the interface ($k_y=0$), the \textit{k}-vector does not change its direction and only increases its magnitude. It results in the situation where the original DE mode is no longer conserved -- the \textit{k}-vector is not perpendicular to magnetization. The group velocity direction and magnitude also change. The change in the group velocity direction is 28°, which is almost double than the change of the magnetization direction.

In Fig.~\ref{fig:dispersions}(b) we analyze the situation where the magnetization in the left region is tilted by 18°, and the \textit{k}-vector is tilted (53° from the $x$-axis) in a way that the curvature of the spin wave dispersion isofrequency line is zero and the group velocity direction is stationary around neighboring \textit{k}-vectors\cite{wartelle23, veerakumar06} (caustic direction). The spin wave then propagates towards the interface with perpendicular group velocity. Again, the \textit{k}-vector component parallel to the interface must be conserved, thus the \textit{k}-vector changes its magnitude and direction. In this situation, the change is not so dramatic as in the previous case -- the \textit{k}-vector direction rotates by 9°, whereas the group velocity is turned by 14°. Even though the \textit{k}-vector is not fully conserved, the selected caustic direction (zero curvature of the isofrequency line) allows to compensate for this mismatch and the group velocity could still follow the turn. 

In order to keep the spin-wave wavelength (\textit{k}-vector magnitude) unchanged upon refraction, we must pay attention to the orientation of the interface with respect to the \textit{k}-vector of the incoming wave [see Fig.~\ref{fig:dispersions}(c)]. If the interface is tilted in a way that it creates a symmetry axis between the incoming and refracted \textit{k}-vectors, then the conservation of the \textit{k}-vector component parallel to the interface is fulfilled and at the same time the \textit{k}-vector magnitude is conserved. Only its direction changes, exactly following the magnetization vector's direction. By stacking multiple interfaces, it is possible to achieve unlimited control over the spin wave propagation direction while conserving its wavelength.

\begin{figure}
\includegraphics[width=85mm]{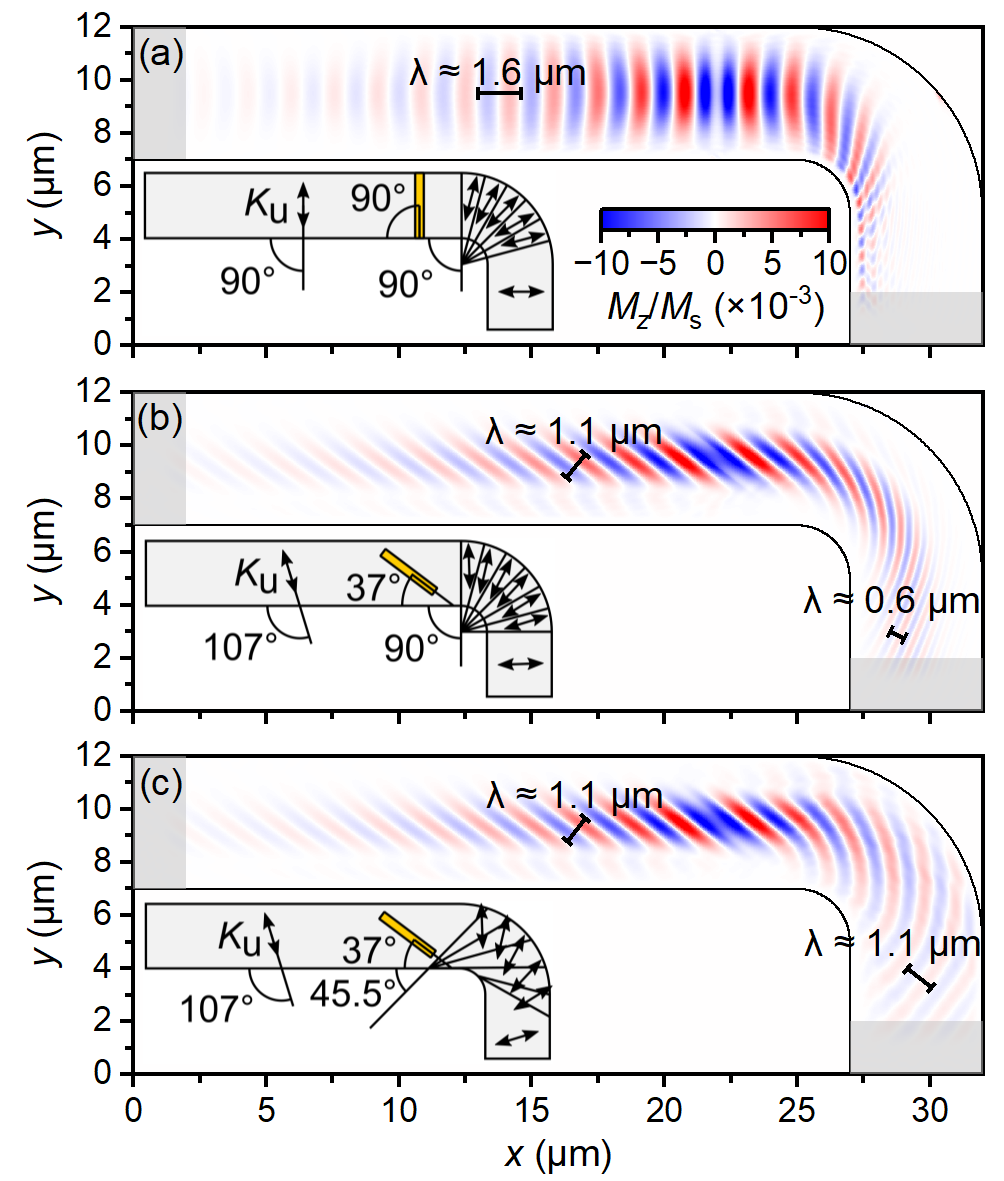}
\caption{\label{fig:simulations} Micromagnetic simulations of spin wave propagation in waveguides with 90° bend with different magnetic configurations. The details of the static magnetic configuration are shown in insets. The gray rectangles mark damped regions. (a)~Waveguide (WG1) with magnetization stabilized perpendicularly to the long axis of the waveguide. (b)~Waveguide (WG2) with tilted magnetization and interfaces between sections always perpendicular to the long axis of the waveguide. (c)~Waveguide (WG3) with tilted magnetization and interfaces oriented in a way that they allow for conservation of the \textit{k}-vector magnitude upon refraction.}
\end{figure}

To test this concept, we performed micromagnetic simulations using MuMax3 micromagnetic solver\cite{vansteenkiste14}, using the same material parameters as those for the calculation of the spin wave refraction in Fig.~\ref{fig:dispersions}. The damping parameter was set to $\alpha = 0.005$ and the cell size was set to 20×20×10$\,\mathrm{nm}^3$. Based on the three cases of refraction in Fig.~\ref{fig:dispersions}, we created three types of 5$\,$µm wide and 37$\,$µm long magnonic waveguides with a 90° bend at one end (Fig.~\ref{fig:simulations}). The spin waves in these waveguides were excited by applying a sinusoidal, magnetic field with the frequency $f = 6\,$GHz and the amplitude $B_{\mathrm{rf}} = 1\,$mT, in 500$\,$nm wide rectangular regions in their center [see yellow rectangles in insets in Fig.~\ref{fig:simulations}, panels (a--c)]. The excitation field always pointed in-plane, perpendicularly to the long edge of the rectangular excitation region. The angle of the rectangular region then defined the direction of the \textit{k}-vector of excited spin waves. To excite well-defined beams, the amplitude of the excitation field was decaying towards the edges as a Gaussian function. Note that in Fig.~\ref{fig:simulations}(a) the beam is wider due to the larger projection of the excitation region's length across the waveguide. To prevent unwanted reflections from the waveguides’ ends, we created regions with quadratically\cite{venkat18} increasing damping ($\alpha \to 1$) in the last 2 micrometers of each end of the waveguide. The magnetic configuration of each waveguide was set correspondingly with the three cases discussed earlier (see Fig.~\ref{fig:dispersions}). To force the magnetization to remain in the desired direction, we set the uniaxial anisotropy strength to $K_{\mathrm{u}}=8.92\,\mathrm{kJ/m^3}$ and varied its direction in different sections of the waveguide [see the insets in Fig.~\ref{fig:simulations}, panels (a--c)]. Note that we had to slightly increase the value of the uniaxial anisotropy and change the angle of its axis (compared to the analytical model in Fig. 1) to compensate for the demagnetizing field from the finite geometry of the waveguide.

The first waveguide (WG1), shown in Fig.~\ref{fig:simulations}(a) represents the situation, where the magnetization is always perpendicular to the long axis of the waveguide and the spin waves propagate in the straight section of the waveguide in DE geometry. When the wave reaches the turn and the first rotated section of the waveguide, it must accommodate the new dispersion and changes its \textit{k}-vector. Also, its group velocity changes by a larger angle than the angle of rotation of the magnetization [see Fig.~\ref{fig:dispersions}(a)] and the wave is steered towards one side of the waveguide. This overturning effect practically blocks the propagation of the spin wave through the 90° bend and is also visible in less sharp bends\cite{vogt12}.  

In the second waveguide (WG2), the magnetization is rotated by 18° and the \textit{k}-vector by 53° compared to the first waveguide; see inset in Fig.~\ref{fig:simulations}(b) and compare it with Fig.~\ref{fig:dispersions}(b). The spin wave group velocity still points along the long axis of the waveguide. Note, that the group velocity in this configuration is higher ($v_{\mathrm{g,tilted}} = 2.8\,$µm/ns), compared to the group velocity of the wave in DE geometry ($v_{\mathrm{g,DE}} = 2.4\,$µm/ns) which leads to larger decay length of the wave in the second waveguide. When the wave reaches the turn and the first rotated section, its \textit{k}-vector still needs to compensate for the rotated dispersion while fulfilling the condition of conservation of its components parallel to the interface [see Fig.~\ref{fig:dispersions}(b)]. Even though the wave's mode is not conserved and the wavelength shortens as the wave propagates through the turn, the rotation of the group velocity aligns with the rotation of the magnetization, and the wave can nicely propagate through the turn [Fig.~\ref{fig:simulations}(b)].

The third waveguide (WG3) is configured in a way that the magnitude of the spin-wave \textit{k}-vector should be conserved through the whole turn. Now, not only the magnetization is tilted, but also the interfaces between the individual sections of the waveguide are rotated [see inset in Fig.~\ref{fig:simulations}(c)] in a way, that the magnitude of the \textit{k}-vector remains constant upon transition through the interface [see Fig.~\ref{fig:dispersions}(c)]. In this type of waveguide, the spin wave propagates nicely through the turn and its wavelength is kept constant. Unfortunately, the waves originating in different positions across the width of the waveguide refract at different distances from the point of their excitation which leads to a considerable broadening of the spin-wave beam [Fig.~\ref{fig:simulations}(c)].

\begin{figure}
\includegraphics[width=85mm]{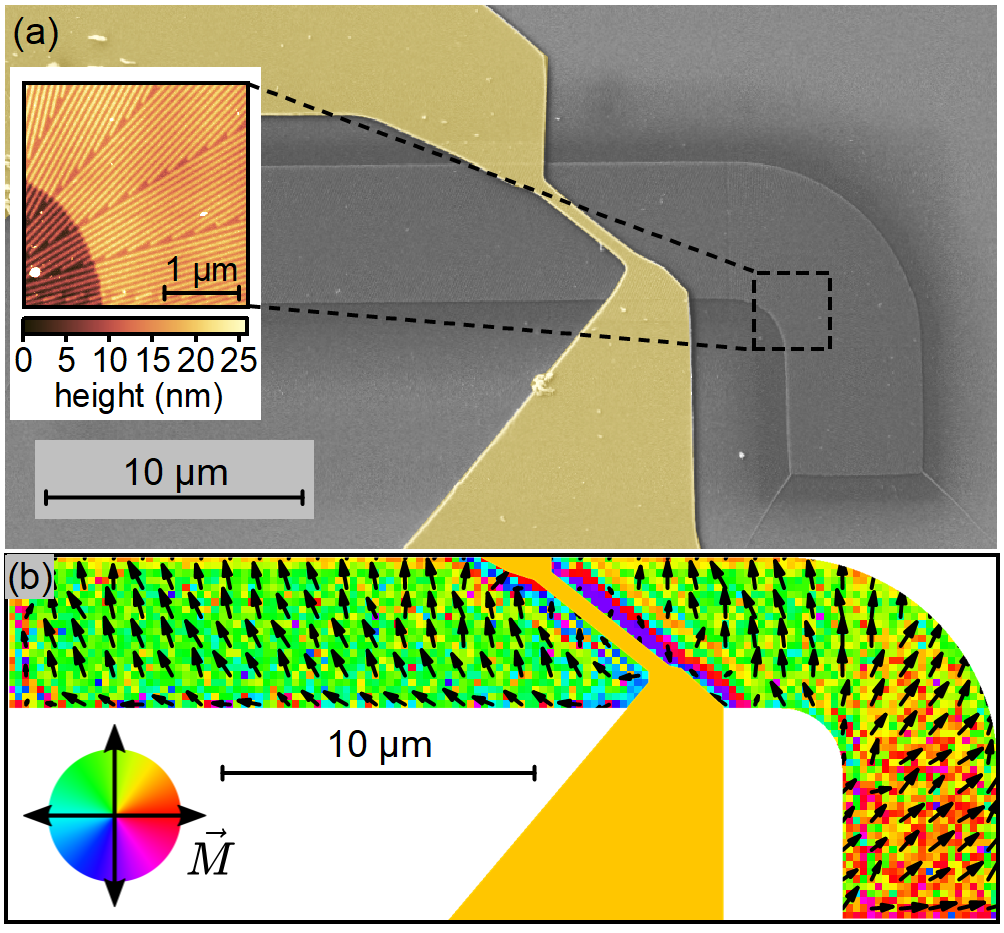}
\caption{\label{fig:staticCharacterization} Static characterization of the fabricated WG2 waveguide. (a)~SEM micrograph of the corrugated waveguide with microwave antenna on top (in ochre color). The inset shows an atomic force microscope image of the corrugation in the bend. (b)~Vectorial reconstruction of the zero-field magnetization $\vec{M}$ acquired by component-selective wide-field Kerr microscope.}
\end{figure}

To experimentally realize the previously discussed regimes of spin wave propagation through 90° bends, we exploited corrugation-induced uniaxial magnetic anisotropy to locally rotate the in-plane magnetization vector into the desired direction\cite{turcan21}, and fabricated corrugated Permalloy (Py, Ni$_{81}$Fe$_{19}$) waveguides segmented into regions with different directions of corrugation [see inset in Fig.~\ref{fig:staticCharacterization}(a)]. The waveguides were prepared by combination of focused electron beam-induced deposition, electron beam lithography, e-beam evaporation, and lift-off techniques on GaAs substrate using procedure described in\cite{turcan21}. To create sinusoidal surface patterns, we scanned the 30$\,$kV electron beam in a series of single lines separated by a distance of 80$\,$nm while introducing the SiO$\mathrm{_2}$ precursor into the microscope vacuum chamber. We set the number of scans to 5,000, which resulted in the corrugation amplitude of 11$\,$nm. Then, in the next step, we defined the required shapes of the waveguides on top of the pre-prepared sinusoidal structure by e-beam lithography, and deposited the 10$\,$nm Py layer in the e-beam evaporator. For spin-wave excitation, we patterned 0.5$\,$µm wide, 5/25/5/85/10$\,$nm thick (Ti/SiO$_2$/Ti/Cu/Au) microwave antennas oriented in a way, that they excited spin waves with the desired \textit{k}-vector direction. The final structure consisting of the waveguide WG2 with the excitation antenna is shown in Fig.~\ref{fig:staticCharacterization}.

The final amplitude of corrugation of each waveguide was checked by atomic force microscopy [see inset in Fig.~\ref{fig:staticCharacterization}(a), waveguide WG2]. Further, to confirm the final magnetization configuration of the waveguide, we checked the final zero-field state of the waveguide using component-selective wide-field Kerr microscope\cite{soldatov17, mccord15}. We confirmed that the imprinted anisotropy was sufficient to force magnetization to follow the corrugation direction and the single-domain state of the waveguides, see Fig.~\ref{fig:staticCharacterization}(b).

\begin{figure*}
\includegraphics{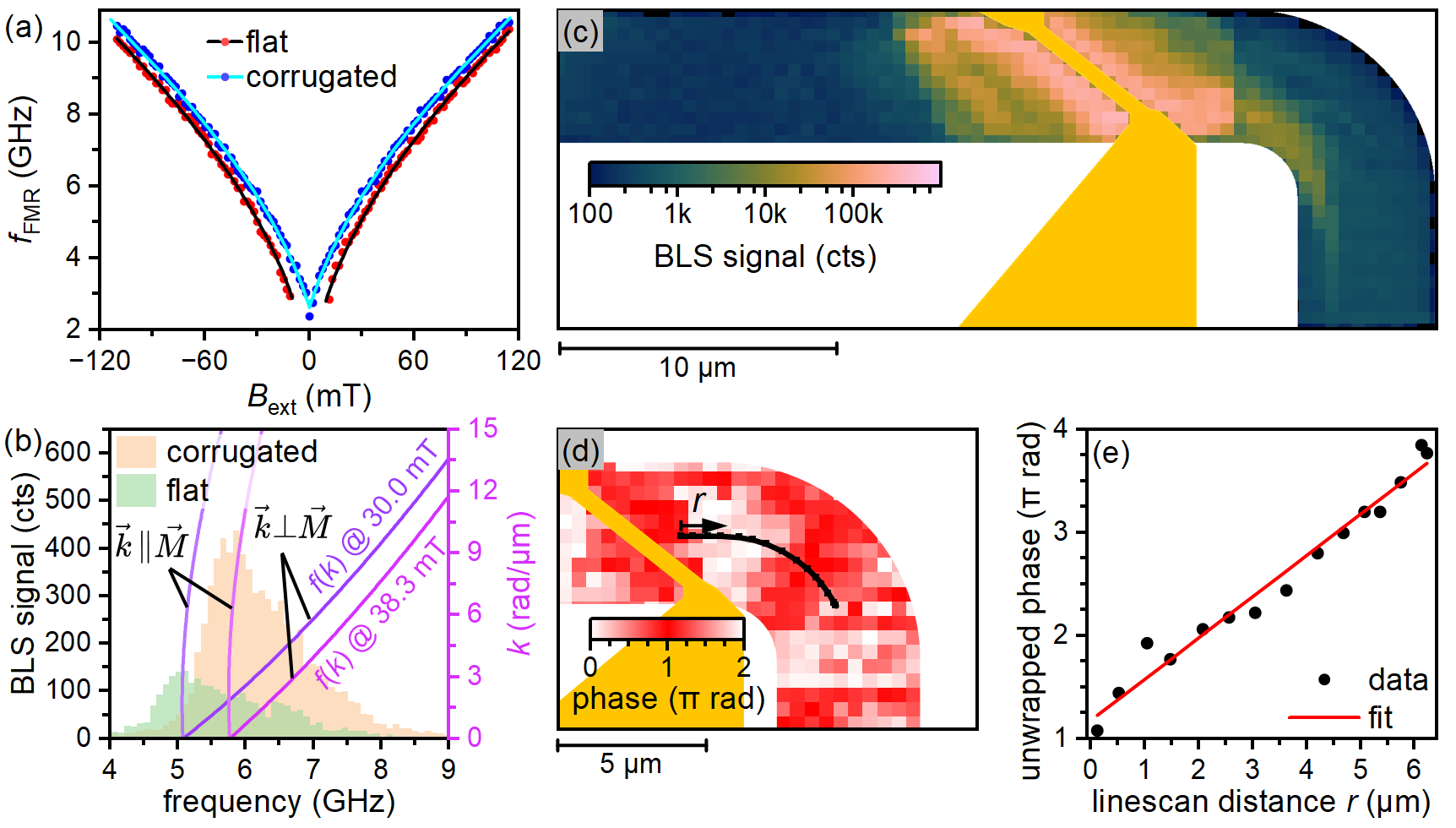}
\caption{\label{fig:BLSmeasurements} Dynamic characterization of the waveguide turns by Brillouin light scattering microscope. (a)~Field-dependent ferromagnetic resonance frequencies obtained from thermal BLS spectra for corrugated (WG2) and reference flat waveguides. The spectra for corrugated [flat] waveguide were fit by Kittel formula with the following parameters: $M_{\mathrm{s}} = (751\pm10)\,$kA/m, $\mu_0H_{\mathrm{a}} = (8.3 \pm 0.3)\,$mT [$M_{\mathrm{s}} = (761\pm4)\,$kA/m, $\mu_0H_{\mathrm{a}} = (0.04\pm0.23)$\,mT]. The value of $\gamma/2\pi = 29.5\,$GHz/T was fixed in both fits\cite{wojewoda23a}. (b)~BLS spectra measured in the external field of 30\,mT, together with analytical calculations of the corresponding spin wave band. (c)~Spin wave propagation through the bend of WG2 waveguide. The sharp decrease of the spin wave intensity at the start of the bend is caused by imperfect stitching of the two neighboring regions of corrugation. (d) Phase-resolved BLS map in the turn of WG3 waveguide with propagating $f=3.5\,$GHz spin wave. Black line shows the position of the linescan shown in panel (e). Linear evolution of the unwrapped phase through the turn indicates conservation of the \textit{k}-vector magnitude. The slope of the phase $k_{v_{\mathrm{g}}} = (1.26\pm0.06)\,$rad/µm here represents a projection of the \textit{k}-vector into the group velocity direction.}
\end{figure*}

After checking the static configuration of the magnetization, we transferred the sample with waveguides WG2 and WG3 to our BLS microscope and performed a series of dynamic experiments. In the first experiment, we measured thermal BLS spectra at different values of the external field for the corrugated, and for the reference flat waveguide [Fig.~\ref{fig:BLSmeasurements}(a, b)]. From fitting the FMR frequency with Kittel formula\cite{hache20, kittel48} we obtained the value of corrugation-induced anisotropy field $\mu_0H_{\mathrm{a}} = (8.3 \pm 0.3)\,$mT, whereas in the flat reference waveguide, the (shape) anisotropy field was negligible. Fig.~\ref{fig:BLSmeasurements}(b) shows the BLS thermal spectra acquired for the corrugated and reference flat waveguides in the external magnetic field of 30\,mT, together with analytically calculated dispersion relations\cite{kalinikos86, kalinikos90, wojewoda23b} for two values of the effective magnetic field. By projecting the frequency axis to the \textit{k}-vector axis using the calculated dispersion relation (and taking into account broadening by finite lifetime of spin waves and transmission function of tandem Fabry-Perot interferometer\cite{lindsay81}), we can estimate that we can detect spin waves with \textit{k}-vectors up to 7\,rad/µm. Note that the BLS signal from the corrugated waveguide is approx. three times higher than the signal from the flat waveguide. This is caused by the fact that the elliptical precession of the magnetic moments follows the corrugated surface and in the regions between the hills and valleys of the sine function the precession ellipse is tilted. This increases the projection of the dynamic part of the magnetization into the direction perpendicular to the sample plane, which leads to the increase of the BLS signal (which is, in our configuration, mainly sensitive to out-of-plane dynamic component of the magnetization \cite{cottam76, wojewoda23a}).

After the thermal measurement, we connected the antenna to an RF generator and performed a series of experiments with propagating spin waves at zero external magnetic field. The excitation frequency was set to $f = 3.5\,$GHz to maximize the BLS signal after the bend. Fig.~\ref{fig:BLSmeasurements}(c) shows BLS intensity map of 3.5\,GHz spin wave propagating through the 90° bend of the WG2 waveguide. It can be clearly seen that the spin wave propagates dominantly along the long axis of the waveguide and through the turn (even when the excitation antenna is tilted).

In the last experiment we performed a phase-resolved measurement\cite{wojewoda23c, flajsman22, serga06} of the spin wave propagating through the turn of WG3 waveguide. In Fig.~\ref{fig:BLSmeasurements}(d) we can see a spin-wave phase map together with extracted linescan through the center of the waveguide [Fig.~\ref{fig:BLSmeasurements}(e)]. Linear evolution of the phase in the linescan confirms, that the mode of the spin wave does not change through the turn. The slope of the phase $k_{v_{\mathrm{g}}} = (1.26\pm0.06)\,$rad/µm here represents a projection of the \textit{k}-vector into the group velocity direction. To obtain the real value of the \textit{k}-vector we need to take into account the angle between $\vec{k}$ and $\vec{v}_{\mathrm{g}}$ which, for our system, is approx. 53°. The value of $k$ is then approx. 2\,rad/µm, which corresponds to the spin wave wavelength $\lambda=3.14\,$µm.

In conclusion, we have shown that it is possible to smoothly and coherently steer dipolar spin waves in in-plane magnetized waveguides through 90° turn, despite their anisotropic dispersion relation. To achieve undisturbed propagation, a specific magnetization configuration of the waveguide, taking into account the conservation of the \textit{k}-vector component parallel to the interface, needs to be prepared. This condition prohibits turning the spin waves in the most common geometries ($\vec{k}\perp \vec{M}$, Damon-Eshbach and $\vec{k}\parallel \vec{M}$, backward volume) without wavefront distortion, i.e., without oversteering and change of the wavelength through the turn. To achieve a single-mode turn, the spin wave \textit{k}-vector must point in specific direction, different from the direction of propagation (group velocity). The optimum working point seems to be the caustic direction, i.e. the direction where the curvature of the spin wave dispersion is zero and the group velocity direction is stationary around some \textit{k}-vectors\cite{wartelle23, veerakumar06}. The major challenge in this type of turn lies in the realization of the proper magnetization landscape. In our work, we achieved the required landscape through laborious 3D nanopatterning of the individual sections of the waveguides. Although this technique can be further perfected, additional miniaturization can be a challenge as the demagnetizing field of narrower, sub-micrometer waveguides can overcome the corrugation-induced anisotropy and prevent the magnetization to be stabilized in the desired direction. Here, alternative approaches controlling directly magnetocrystalline anisotropy\cite{urbanek18, flajsman20, wojewoda20} could also be exploited. The findings presented in this work should be considered in any magnonic circuit design dealing with anisotropic dispersion and spin wave turns.

\begin{acknowledgments}
This work was supported by the Grant Agency of the Czech Republic, project no. 23-04120L. CzechNanoLab project LM2023051 is acknowledged for the financial support of the measurements and sample fabrication at CEITEC Nano Research Infrastructure. Specific research projects IDs CEITEC VUT-J-23-8435 and CEITEC VUT-J-23-8411 are also acknowledged. O.W. was supported by Brno PhD talent scholarship.
\end{acknowledgments}

\section*{AUTHOR DECLARATIONS}
\subsection*{Conflict of Interest}
The authors have no conflicts to disclose.
\subsection*{Author Contributions}
\textbf{Jan Klíma} Conceptualization (equal), Investigation (lead), Visualization, Resources (lead), Writing – review \& editing (equal); \textbf{Ondřej Wojewoda} Investigation (equal), Methodology, Writing – review \& editing (equal); \textbf{Václav Roučka} Conceptualization (equal), Writing – review \& editing (equal); \textbf{Tomáš Molnár} Investigation (supporting), Writing – review \& editing (equal); \textbf{Jakub Holobrádek} Resources (supporting), Writing – review \& editing (equal); \textbf{Michal Urbánek} Conceptualization (equal), Funding acquisition, Project administration, Supervision, Writing – review \& editing (lead)

\section*{DATA AVAILABILITY}
The data that support the findings of this study are available from the corresponding author upon reasonable request.

\section*{REFERENCES}
\bibliography{aipsamp}

\end{document}